\documentclass[times]{TRR}

\usepackage{graphicx}
\usepackage{tabularx, booktabs}
\usepackage{tabulary}
\usepackage{multirow}
\usepackage{subcaption}
\usepackage{algorithm}
\usepackage[noend]{algpseudocode}
\MakeRobust{\Call}
\DeclareMathOperator*{\argmin}{arg\,min}
\usepackage[colorinlistoftodos]{todonotes}

\usepackage[colorlinks=true,linkcolor=blue,citecolor=blue]{hyperref}

\usepackage{tikz,xcolor}

\begin{document}


\runninghead{Sundt et al.}

\title{Heuristics for Customer-focused Ride-pooling Assignment}

\author{Alexander Sundt \affilnum{1}, Qi Luo \affilnum{2}, John Vincent \affilnum{3}, Mehrdad Shahabi \affilnum{3}, and Yafeng Yin \affilnum{1}}

\affiliation{\affilnum{1} Department of Civil and Environmental Engineering, University of Michigan\\
\affilnum{2} Department of Industrial Engineering, Clemson University \\
\affilnum{3} Robotics and Mobility Research, Ford Motor Company}

\corrauth{Yafeng Yin, yafeng@umich.edu}





\begin{abstract}
Ride-pooling has become an important service option offered by ride-hailing platforms as it serves multiple trip requests in a single ride.
By leveraging customer data, connected vehicles, and efficient assignment algorithms, ride-pooling can be a critical instrument to address driver shortages and mitigate the negative externalities of ride-hailing operations.
Recent literature has focused on computationally intensive optimization-based methods that maximize system throughput or minimize vehicle miles.
However, individual customers may experience substantial service quality degradation due to the consequent waiting and detour time.
In contrast, this paper examines heuristic methods for real-time ride-pooling assignments that are highly scalable and easily computable.
We propose a restricted subgraph method and compare it with other existing heuristic and optimization-based matching algorithms using a variety of metrics.
By fusing multiple sources of trip and network data in New York City, we develop a flexible, agent-based simulation platform to test these strategies on different demand levels and examine how they affect both the customer experience and the ride-hailing platform.
Our results find a trade-off among heuristics between throughput and customer matching time.
We show that our proposed ride-pooling strategy maintains system performance while limiting trip delays and improving customer experience.
This work provides insight for policymakers and ride-hailing operators about the performance of simpler heuristics and raises concerns about prioritizing only specific platform metrics without considering service quality.
\end{abstract}

\maketitle



\section{Introduction}
The ridesourcing industry has recently grown tremendously due to the rise of transportation network companies (TNCs) such as Uber, Lyft, and DiDi Chuxing. Already serving billions of passenger trips per year, this industry has the potential to reshape cities, reduce the need for parking, and fortify the transportation ecosystem  \cite{wang2019ridesourcing,tafreshian2020frontiers}. On-demand mobility services offered by TNCs also improve accessibility for those living in transit deserts with limited mobility choices.

While ridesourcing has revolutionized the ground passenger transportation market, it has also come with undesirable consequences. The increase in ridesourcing trips has drawn riders from public transit and increased the congestion in urban areas \cite{henao2019impact, hall2018uber, luo2019dynamic}. In San Francisco alone, the County Transportation Authority (SFCTA) found that TNCs were responsible for more than half of the 60 \% increase in traffic congestion between 2010 and 2016 \cite{hawkins_2019}. The negative congestion externality of ridesourcing is primarily due to the increase in vehicular traffic demand. The convenience and flexibility of these services have both induced travel and encouraged a modal shift. The shift increases vehicular traffic demand if the original modes of transportation are non-motorized such as walking or biking. It also yields additional traffic demand if the original modes have higher occupancy, such as public transit and private driving. Compared to these modes, the average occupancy of ridesourcing vehicles is much lower because of the massive amount of empty miles traveled in the system. These are distances that ridesourcing vehicles travel when searching for or picking up a request, and would not happen if the trip was made in a personal vehicle. In summary, ridesourcing services have yielded an increase in vehicle miles traveled (VMT), causing congestion on city streets. The inefficient operations of these ridesourcing platforms can have severe environmental impacts due to increased energy consumption and emissions. 

The potential for vast growth in the ridesourcing industry must be managed efficiently without slowing cities' traffic to a halt. Both planners and the services themselves have considered strategies to govern this. Transportation authorities may implement external regulations, such as capping the number of TNC vehicles or congestion pricing to stimulate behavioral changes in the ridesourcing market \cite{luo2019dynamic, erhardt2019transportation}. TNCs, internally motivated to enhance system performance by reducing the empty miles  \cite{braverman2019empty}, have also implemented tactical policies, such as introducing meeting points so that passengers can walk and shorten the pickup distance \cite{stiglic2015benefits}. In addition, encouraging shared rides (termed as ``ride-pooling'') when possible is another effective way for TNCs to increase the utilization of ridesourcing vehicles. However, the benefit of ride-pooling is dependent on using efficient ride-pooling assignment algorithms. In each assignment, we need to allocate multiple rides with compatible routes to one available vehicle and the assigned vehicle needs to determine the best route to pick up and drop off these rides. Thus, it is critical to study how to dispatch vehicles for ride-pooling services efficiently to offset the excess VMT and limit the negative externalities of conventional ridesourcing services.

It is well-known that the trip-vehicle assignment problem for ride-pooling is notoriously difficult, and most solutions involve computationally-heavy optimizations. Previous papers have quantified the benefit of ride-pooling by solving large-scale multi-vehicle dial-a-ride problems or approximate dynamic programming \cite{santi2014quantifying, Alonso-Mora462, yu2019integrated, simonetto2019real}. Note that some of these papers assume an offline setting where trip requests are known in advance. 

Solving ride-pooling assignment to global optimality for real-time operations is even more challenging, if not impossible.  More practical alternative approaches include online approximation algorithms, reinforcement learning, and heuristics (including metaheuristics). Online approximation algorithms can provide provable guarantees on the computed solution with respect to the optimum \cite{ashlagi2018maximum, bei2018algorithms} in the assignment of trip requests arriving incrementally over time.  However, these settings are restricted to certain instances and are unrealistic because existing studies focus on analyzing the performance of these approximation algorithms in the worst case. These algorithms' performance on real-time spatial data from cities show various negative results. For example, \citep{tong2016online} discovered that theoretically competitive algorithms might have worse average-case performance than a simple greedy assignment algorithm.  Reinforcement learning is a purely data-driven approach \cite{jindal2018optimizing} that has been used extensively in industry. Nevertheless, reinforcement learning may suffer from lacking adequate real-world data in early adoption and new areas (leading to poorly trained models) or may be lead to unpredictable behavior in unforeseen scenarios. This work aims to examine potential rule-based heuristic vehicle-trip assignment methods. These heuristic methods are not computationally intensive, easily scalable, and prioritize sharing rides over single rides.

Evaluating heuristics for ride-pooling can be difficult and subjective due to the lack of well-balanced performance measurements.  Most heuristic methods in the prior work fall into the ``platform-focused'' catalogue \cite{pelzer2015partition, tong2018unified}. The objective is serving the most trips in the fewest VMT. Nevertheless, implementing poorly designed heuristics may significantly increase customers' travel time compared with the conventional single-ride service. The platform must then compensate the customers by offering a discount. This work proposes a new ``customer-focused''  heuristic method to reduce the service degradation due to ride-pooling.  This method guarantees the detours from the customers' trip plans are restricted, hence called the \emph{restricted subgraph method}. This novel heuristic method could potentially achieve similar performance to optimization-based methods while simplifying calculations.

\subsection{Main contributions}
The main contributions of this work include:
\begin{enumerate}
    \item Suggest a set of well-balanced performance measures for evaluating ride-pooling systems.
    \item Propose a new, simple but effective heuristic and examine a family of heuristic methods for ride-to-ride and ride-to-vehicle assignment that improves the customers' ride-pooling experience.
    \item Build a data-driven agent-based simulator upon which to evaluate and compare different heuristics. 
\end{enumerate}

The remainder of the paper is organized as follows. In Section \ref{sec2}, we review the relevant literature on offline and real-time ride-pooling assignment algorithms. In Section \ref{sec3}, we suggest a comprehensive set of metrics to evaluate the performance of ridesourcing assignment methods.  In Section \ref{sec:heuristic}, we propose a customer-focused heuristic based on evaluating restricted subgraphs of trip requests. In Section \ref{sec4}, we describe how the numerical experiments are conducted and discuss the numerical results. We draw the conclusion in Section \ref{sec5}.

\section{Literature review} \label{sec2}
There have been burgeoning studies to investigate the ride-pooling assignment problem. The goal is to dispatch ride-hailing vehicles to enable customers to share rides with others going in the same direction in order to offset travel costs with multiple passengers in the vehicle. Much of the prior optimization-based work assumes that trip requests are known ahead of time, and therefore the proposed approach is inherently offline. Others instead receive knowledge of trip requests in batches and perform computationally demanding optimizations to determine routes and matches. On the other hand, those for pure online implementation with unknown demand were designed against the worst case. In contrast, this work seeks to approach the middle ground for easily computable, real-time ride-pooling assignment methods. 


Previous studies assuming that the platform has partial or full knowledge of travel demand focused on developing centralized, optimization-heavy approaches to improve the overall system performance. 
The resulting benefit of ride-pooling is over-optimistic, as real-time ridesourcing operations often do not know about future requests or have time to perform intensive calculations. \citep{Alonso-Mora462} has indicated the substantial potential of high-capacity ride-pooling. This alternative on-demand transportation system can serve 98\% of the demand with 15\% of the fleet size when using a fleet of small busses with a capacity of 10 people. They used the idea of ``shareability graphs'' (vehicle-trip-request graphs) to optimize among potential matches of multiple trips and vehicles. This approach requires solving the Travelling Salesman Problem repeatedly for each potential pairing of trips to check if constraints regarding pick up time and delay are satisfied and then requires solving an integer program to optimize the miles traveled. Both of these problems are highly computationally intensive and can scale exponentially with the number of requests. \citep{simonetto2019real} improved the computational efficiency of the algorithm as a linear assignment problem and distributing the computation to multiple platforms. The results achieved a similar quality of service and ran up to four times faster than in the prior work. Since the assignment algorithm's performance varies significantly when the information about the next demand is unknown, this stream of literature can be treated as an upper bound on any assignment policies in real-time/online ride-pooling systems. 

On the other hand, there is a growing body of approximation algorithm literature that is derived from a worst-case analysis. These analyses are over-conservative because they examine the performance of algorithms when requests arrive in an adversarial order, which is designed to elicit poor results. Regarding the conventional single-ride vehicle assignment (i.e., each vehicle is matched with one trip request), \citep{tong2016online} revealed that a simple greedy assignment algorithm, in which requests are matched on arrival to the closest driver, outperformed randomized online matching algorithms such as the permutation algorithm and the hierarchically separated tree algorithm. A potential reason is because those worst cases are rare in practice. 
\citep{ashlagi2018maximum} proposed a passenger-to-passenger matching algorithm in ride-pooling assignment systems where each customer has either a constant or a random deadline. Their algorithm, based on the adversarial arrival order assumption, achieved $1/4$-competitive for fixed deadline and $1/8$-competitive when the demand process is memoryless. \citep{azar2017online} initiated the studies of online matching with delays. This model captured the trade-off between the market thickness and information in the dynamic matching systems. In the context of ride-pooling, the longer customers stay unmatched, the higher probability that a better match can be found. 
For those works, ridesourcing platforms may pose the same question as to the single-ride case -- how do they perform on the real-world spatial data? This paper aims to cast light on this question.

Work in the area of real-time operation of ride-pooling systems has become more prevalent in recent years. The vehicle-trip assignment problem in ridesourcing is a variation of the dial-a-ride problem, i.e., designing efficient vehicle routes to serve the users' pickup and delivery requests between origins and destinations. 
Easily computable heuristic methods to solve the vehicle-trip assignment in ridesourcing were examined by  \citep{HYLAND2018278}. This showed the potential of different methods of prioritization and matching, as well as considering varying scenarios of driver statuses. However, \citep{HYLAND2018278} only examined the single-trip-per-ride case and did not consider the potential for pooling or sharing. \citep{herminghaus2019mean} analyzed the efficiency of ride-pooling by using a mean-field approach in different urban settings. The ride-pooling market is characterized by aggregate variables such as the distribution of demand, the average delay of ride-pooling, and the traffic network structure. The model explained why mobility-on-demand is more competent in dense urban areas and revealed a break-even point for its deployment in urban areas.

\section{Performance measures of ride-pooling systems} \label{sec3}
Designing real-time ride-pooling heuristics starts with a clear goal in mind, but a consistent and objective measure of ride-pooling effectiveness is lacking in this avenue of research. The following performance measures are widely used to guide the design and the operations of ridesourcing services:
\begin{enumerate}
    \item Maximize the utilization of ridesourcing vehicles \cite{cramer2016disruptive}.
    \item Maximize average vehicle occupancy  \cite{di2017ridesharing}.
    \item Maximize the platform's throughput \cite{MASOUD2017218}.
\end{enumerate}

However, these measures do not fully reflect the saved VMT due to ride-pooling, which is the main benefit of its deployment. Utilization specifically does not distinguish between time spent detouring (increased VMT) and time spent en-route with multiple passengers (decreased VMT). Designing heuristics around these measures alone may also lead to unintended consequences. For example, if an algorithm cherry-picks short-haul trips, the system's overall throughput can significantly outperform other algorithms. Alternatively, if vehicles stay occupied when picking up new passengers, the vehicle utilization will appear high. Passengers may not benefit from these resulting ride-pooling algorithms due to being treated unequally or being forced on long detours and increasing VMT. To better evaluate the performance of ride-pooling systems, we propose a family of performance metrics summarized in Table \ref{tab:1}.

\begin{table*}
\centering
\small 
\begin{tabulary}{0.8\textwidth}{l|l|l}
\toprule
\multicolumn{2}{c|}{ {\bf Evaluation metrics} }  & \multicolumn{1}{c}{ {\bf Definition} } \\ \midrule \multirow{4}{*}{\begin{tabular}[c]{@{}l@{}}Platform\\ performance \\ metrics\end{tabular}} & System throughput   & \begin{tabular}[c]{@{}l@{}}Number of passengers delivered to their\\ destinations per unit of time\end{tabular}                                           \\ \cline{2-3} 
    & System efficiency & \begin{tabular}[c]{@{}l@{}}Average passenger-hours served \\ per unit labor hour\end{tabular}                                   \\ \cline{2-3} 
    & Time-based vehicle occupancy         & Time-averaged number of passengers on board                                                                                                \\ \cline{2-3} 
    & Distance-based vehicle occupancy  &  Distance-averaged number of passengers on board\\ \hline                                                 
\multirow{3}{*}{\begin{tabular}[c]{@{}l@{}}Customer\\ performance \\ metrics\end{tabular}} & Matching time      & \begin{tabular}[c]{@{}l@{}}Average time between when a trip request is \\ sent and a driver is dispatched to the trip\end{tabular}                   \\ \cline{2-3} 
    & Pickup time & \begin{tabular}[c]{@{}l@{}}Average time between when a driver \\ is dispatched and the customer is picked up\end{tabular}                   \\ \cline{2-3} 
    & Detour time & \begin{tabular}[c]{@{}l@{}}Average difference of in-vehicle time using \\ ride-pooling relative to that of using \\single-ride service\end{tabular} \\ \bottomrule
\end{tabulary}
\caption{Summary of heuristic evaluation metrics} \label{tab:1}
\end{table*}

Below we elaborate on the system throughput and system efficiency, as other performance metrics in Table \ref{tab:1} are self-explanatory and easily measured. For a study period, say, three hours, we discretize it into smaller time intervals of, e.g., five minutes each. At a given time interval $\tau$, we let  $V_{\tau}$ denote the set of in-service vehicles at the interval and $ K_\tau^v$ denote the set of trips delivered by vehicle $v\in V_{\tau}$ during the time window $[\tau, \tau+ \Delta \tau]$ where $\Delta \tau$ is the length of the moving window, e.g., 10 minutes. 

We first introduce system throughput, which is defined as the number of passengers delivered to their destinations per unit of time. If real-time passenger dropoff data are available, as is the case in simulations in this paper, the throughput $Q_\tau$ can be simply measured by the number of passengers dropped off within the window $[\tau, \tau+ \Delta \tau]$ divided by $\Delta \tau$. Otherwise, it can be approximated by the ratio between the total number of passengers on board and the average in-vehicle time of passengers. As per Little's law, this approximation is exact if the system is in a steady state. Let $O_\tau$ denote the average occupancy of vehicles in $V_\tau$. Mathematically, the throughput can be computed as follows:

\begin{align}
    Q_\tau  = \frac{ |V_\tau|  O_\tau}{\sum_{v\in V_{\tau}} \sum_{k\in K_v^\tau}   t^{\text{in-veh}}_{k} / \sum_{v\in V}|K_v^\tau|} ,
\end{align}
where $t_k^{\text{in-veh} }$ is the in-vehicle trip time experienced by the passenger trip $k\in K_v^\tau$ served by vehicle $v\in V_\tau$ and $\mid \cdot \mid$ represents the size of a set. In the equation, the numerator is the total number of passengers on board while the denominator computes the average in-vehicle trip time. 

System throughput, as defined above, is a good performance measure of the productivity or output of a ridesourcing system. However, it does not reflect the resource, i.e., labor hours, utilized to produce such an output. This is important in a ride-pooling scenario to quantify how much benefit is derived from sharing rides. More importantly, throughput does not reflect the length of the trips served. As previously mentioned, a matching algorithm that maximizes system throughout will tend to cherry-pick to serve short-haul trips. To address these two issues, we propose a system efficiency metric. The adjusted productivity or output at time interval $\tau$ will be the sum of the lengths of all trips (measured by their travel times if being served by single-ride service) delivered during the time window $[\tau, \tau+ \Delta \tau]$. Mathematically, the system efficiency is defined as follows:

\begin{align}
    E_\tau = \frac{\sum_{v\in V_{\tau}}  \sum_{k\in K_v^\tau} t_k}{\overline V_\tau \cdot \Delta \tau},
\end{align}
where $\overline V_\tau $ is the average number of in-service vehicles in $[\tau, \tau+ \Delta \tau]$, and thus the denominator assesses the total labor input. $E_\tau$ indicates the system outcome produced per unit of labor hours through ride-pooling. Note that for systems without ride-pooling, this metric reduces to the well-known system utilization metric, i.e., the percent time-in-service vehicles being occupied. 

Note that $Q_\tau$ and $E_\tau$, as defined, yield time-varying metrics of how a ride-pooing system performs during a study period. For the whole period, we report the averages of these time-varying metrics over all time intervals. Later we conduct statistical tests to determine how significant differences in these averages are over many simulation runs.

\section{Ride-pooling assignment heuristics} \label{sec:heuristic}
In this section, we first introduce several basic versions of ride-pooling assignment methods widely adopted in industry and from literature. Note that most of these methods are platform-focused so customers may experience substantial degradation of service quality due to waiting and detour. Thus, we propose a novel customer-focused heuristic called the \emph{restricted subgraph matching} method. 

\subsection{Preliminaries: benchmark ride-pooling methods}
We present three mainstream ride-pooling heuristics adopted in the industry or from the existing literature \cite{yang2020optimizing}. 

Origin-Destination (O/D) grouping or path clustering is a similarity-based algorithm that combines trips for ride-pooling based on the proximity of their origins and destinations. The advantage of the grouping method is that this type of trip similarity is an intuitive indicator for shareable rides without excessive detour time.  Thus, a ride-hailing vehicle can serve matched trip requests by solving a shortest-path problem. On the other hand, two potential disadvantages emerge. First, ride-pooling efficiency is highly dependent on the thickness and homogeneity of the pooling market. If the market is not thick enough, the pooling will likely fail. Otherwise, passengers have to experience much longer matching time to be successfully pooled. Second, the system needs to determine sequences of pickup and dropoff after collecting all trip requests. The option for vehicles to accommodate new demand once their route is determined is ruled out, missing potential compatible requests in the future.

\begin{figure}
    \centering
    \includegraphics[width = \linewidth]{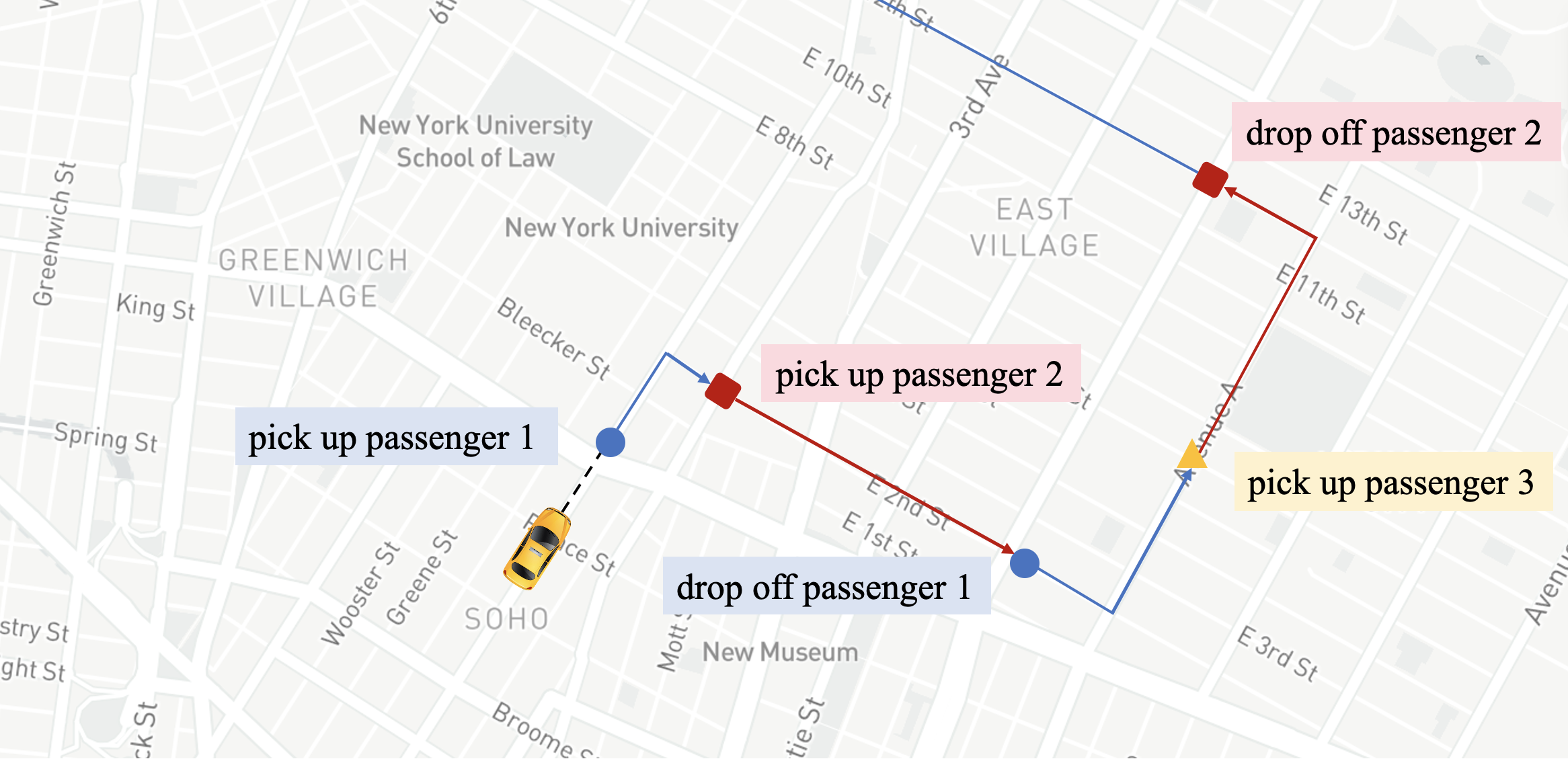}
    \caption{Target occupancy heuristic}
    \label{fig:1}
\end{figure}

A second class of heuristic is occupancy-based. 
The target occupancy heuristic aims to maintain a certain number of passengers per vehicle by alternating pickups and dropoffs. The main idea of the target occupancy heuristic is demonstrated in Figure \ref{fig:1}. As requests are received, they are assigned to the nearest vehicle whose occupancy is below the target. This heuristic should benefit the ridesourcing platform, as it is a greedy algorithm that attempts to maximize the utilization of vehicles. However, since this simplified target occupancy heuristic does not consider requests and vehicle destinations, customers might experience substantial trip delays. To address this issue, we limit the assignment to vehicles within a certain matching radius. The geometric matching radius is effective for limiting both the delay for passengers currently in the vehicle and the pickup time for the assigned request. It is widely used in the ridesourcing industry \cite{yang2020optimizing, zha2018geometric}. 

The final class of the ride-pooling method is bipartite matching or assignment regarding batched requests and vehicles. This method first constructs a bipartite graph in which available drivers (empty cars or cars with vacant seats) and trip requests sit on two sides, respectively. 
Since ride-pooling allows multiple requests to be assigned to the same vehicle, the system is solving a general assignment problem to minimize the total pickup time (or maximize the value of assignments). Efficient approximation algorithms \cite{williamson2011design} can solve the large-scale bipartite matching (conventional) or assignment (ride-pooling) problem. 
The batch size is controlled by how frequently such a bipartite matching is conducted. 
The advantage of this approach is achieving the local optimum for the metrics of interest, in this case total system pickup time, while matching as many requests as possible. 
However, bipartite matching or assignment may cause additional waiting costs if the batch size of matching is extensive, and the detours for requests depend the method is used to determine trip compatibility.

The following counterexamples show how these these heuristics can be arbitrarily bad in specific networks. 

\begin{figure}
    \begin{subfigure}{0.48\linewidth}
    \centering
     \includegraphics[width = \linewidth]{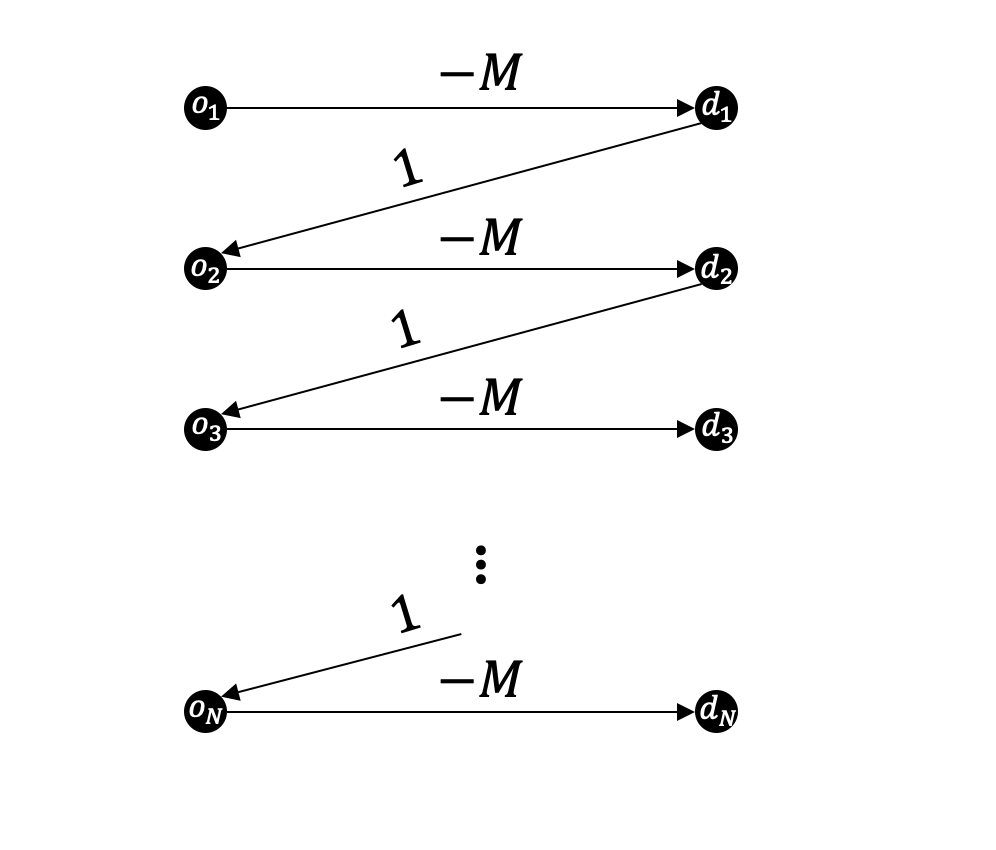}
     \caption{O-D grouping}
    \end{subfigure}
    \begin{subfigure}{0.48\linewidth}
    \centering
     \includegraphics[width = \linewidth]{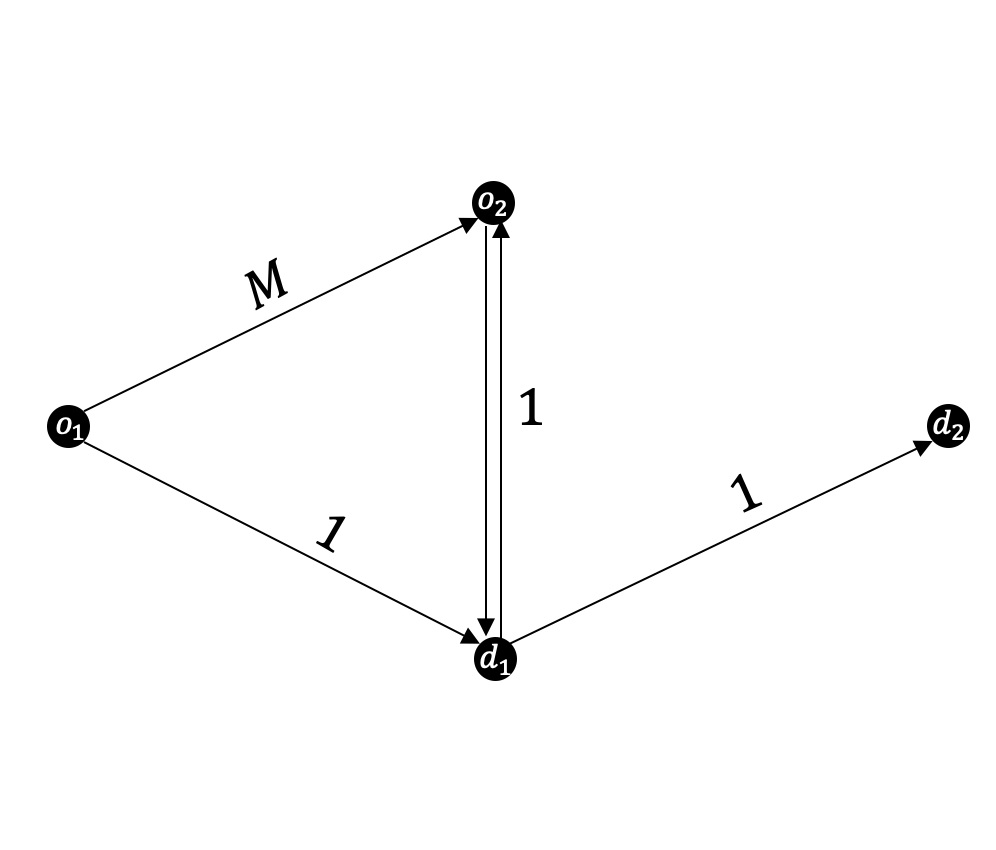}
     \caption{Target occupancy}
    \end{subfigure}
    
    \caption{Counterexamples for benchmark heuristics; $o_i - d_i, i = 1,2, \dots $ are origins and destinations of trip requests on a network, and edge values are travel costs. $M$ is an arbitrarily large positive number. }
    \label{fig:1.5}
\end{figure}

We assume that all trip requests are revealed at time $0$ and the platform operator controls one vehicle with capacity of two to fulfill these demand. The objective is to minimize the total cost of transporting passengers.  In the first counterexample, the operator adopting the O-D grouping algorithm will assign the vehicle to an arbitrary trip to collect a total cost of $-M$, while the obvious optimal path is traveling a sequence of $o_1 \to d_1 \to o_2 \to d_2 \to \dots \to d_N$ with the total cost of $N(1-M)-1$. The optimality ratio is an unbounded $N$.  In the second counterexample, the operator adopting the target occupancy algorithm (with a target of $2$) will choose the path $o_1 \to o_2 \to d_1 \to d_2$ with a total cost of $M+2$. A better path is $o_1 \to d_1 \to o_2 \to d_1 \to d_2$ with a total cost of $4$. The optimality ratio is $(M+2) / 4$, which is also unbounded when we drive $M \to \infty$. Note that in the second counterexample, the first passenger experience extreme detour from the use of target occupancy algorithm. These examples motivate us to develop a family of customer-focused algorithms in what follows.

\subsection{Restricted subgraph method: a customer-focused heuristic}
This section proposes a new customer-focused heuristic built upon the concept of a \emph{restricted subgraph}. 
For each request received and each vehicle in transit, a matchable subgraph is created for the driver and the customer, respectively. 
For a vehicle with passengers, each of these subgraphs contains nodes that can be reached along its route within a reasonable delay; 
Similarly, for a customer seeking a ride-pool, their restricted subgraph contains nodes within an acceptable delay from their initial route.  

The \emph{matchable node} in a subgraph is defined as follows. A request going from node A to node C can reach node B if the sum of the travel time from A to B and the travel time from B to C is less than the original travel time from A to C plus an allowable delay $\delta$ as a function of the original travel time $\delta(tt_{AC})$, i.e., $ tt_{AB} + tt_{BC} \leq tt_{AC} + \delta(tt_{AC})$.

The allowable delay function can be simple constants or reflect more complex, operational considerations. 
In our numerical results, we have chosen $\delta(tt_{AC}) = \sqrt{tt_{AC}}$, as this implies that the acceptable detour time grows with the length of the original trip, but also that for longer trips this delay should not grow at the same rate. For example, a 5-minute detour might be acceptable for a 10-minute trip, but a 30-minute detour would unlikely be acceptable for an hour-long trip.

\begin{figure}
    \centering
    \includegraphics[width = \linewidth]{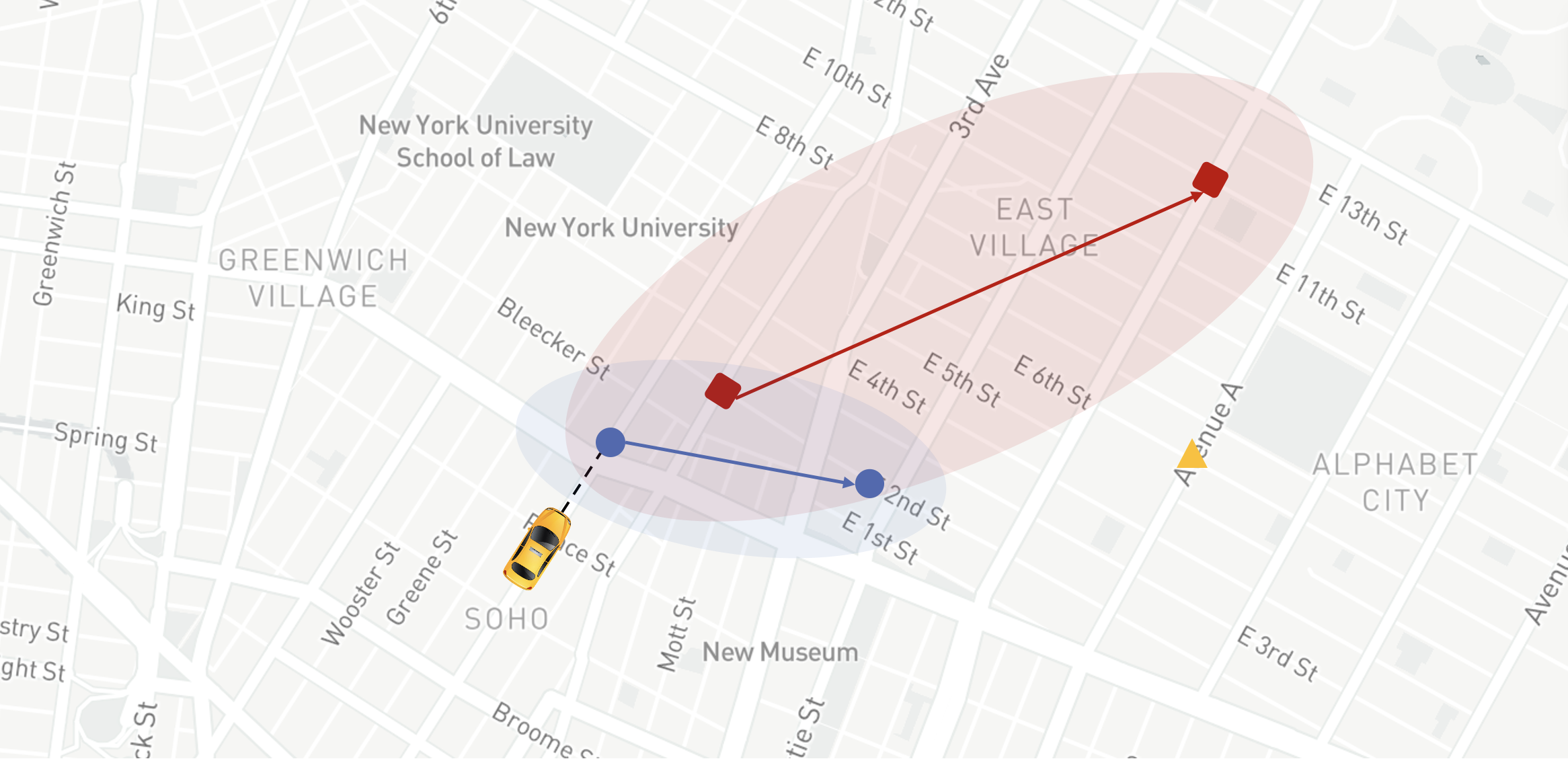}
    \caption{Restricted subgraph heuristic}
    \label{fig:2}
\end{figure}

As can be seen in Figure \ref{fig:2}, constructing this restricted subgraph allows for a concise way of determining whether trips are compatible for pooling. This figure shows that the red origin falls inside the blue trip's subgraph, meaning the red trip can be picked up without too much delay for blue. Similarly, the blue destination falls within the red subgraph, suggesting that the blue trip can be dropped off without too much delay for red. If either trip did not satisfy these subgraph constraints (e.g., the yellow triangle on the map), it would be too far out of the way to yield a reasonable ride-pooling assignment.

Given the travel time ($tt_{AC}$) from A to C, this method considers all nodes within an ellipse of a certain focal distance from the origin and destination. This restricted subgraph method is easy to implement because the node-to-node distances can be pre-calculated. A similar method has been used to assign peer-to-peer carsharing \cite{MASOUD2017218} and check the detour propensity for ride-pooling in the mean-field setting \cite{herminghaus2019mean}.

\begin{algorithm}
\caption{Restricted Subgraph Matching}\label{ressub}
\begin{algorithmic}[1]
\Procedure{match}{$P$, $V_{idle}$, $V_{en-route}$}:
\For{$p \in P$}
    \State $v^* \leftarrow \emptyset$
    \For{$v \in V_{en-route}$}
        \If {$p_o \in G_v$}
            \If {$v_d \in G_{p}$}
                \Comment{Trips overlap}
                \State \Call{assign}{$p$, $v$}
                \State $v^* \leftarrow v$
                \State Break
            \ElsIf {$p_d \in G_v$ and $\Call{dist}{p_d, v_d} < \Call{dist}{p_o, v_d}$}
                \Comment{Passenger trip is a}
                \State \Call{assign}{$p$, $v$} \Comment{subset of vehicle trip}
                \State $v^* \leftarrow v$
                \State Break
            \EndIf
        \EndIf
    \EndFor
    \If {$v^* == \emptyset$ and $|V_{idle}| > 0$}
        \Comment{$p$ can be assigned to idle vehicle}
        \State $v^i \leftarrow \argmin_{v\in V_idle} \Call{dist}{p_o, v_o}$
        \State \Call{assign}{$p$,$v^i$}
        \State $V_{idle} \leftarrow V_{idle} \setminus v^i$
        \Comment{Remove vehicle from idle set}
    \Else
        \State $V_{en-route} \leftarrow V_{en-route} \setminus v^*$ \Comment{Remove vehicle from en-route set}
    \EndIf
\EndFor
\EndProcedure
\end{algorithmic}
\end{algorithm}

Once the subgraphs are created, finding matches is a simple membership test. If a passenger, $p$, can share with a vehicle, $v$, carrying a different passenger en route to their destination, it must first be true that the vehicle can pick up $p$ without delaying $v$ too much. So $p$'s origin, $p_o$, must be in $v$'s subgraph, $G_v$. Additionally, one of the two following cases must be true: $p$'s destination, $p_d$, must be in $G_v$ OR $v$'s destination, $v_d$ must be in $p$'s subgraph, $G_p$. The first case corresponds to the scenario where passenger $p$'s trip is a subset of the vehicle's trip, which means $p$ will be picked up and dropped off before it reaches its original destination. The second case corresponds to when the two trips overlap. After passenger $p$ is picked up, the vehicle will travel to its original destination, drop that passenger off, and travel to passenger $p$'s destination. In our experiments with real-world spatial data, this proposed heuristic has performed well in maximizing the systems' throughput while keeping detours small. 

In addition to the restricted subgraph heuristics, all the other pooling methods tested in this paper are summarized in Table \ref{tab:2}. 
\begin{table*}
\begin{center}
\small
\begin{tabulary}{\textwidth}{l|l|l|l|l}
\toprule
\multicolumn{2}{c|}{{\bf \small Heuristics}}                                                                           & \multicolumn{1}{c|}{ {\bf \small Description} }  &  \multicolumn{1}{c|}{{\bf \small  Pros} } & \multicolumn{1}{c}{ {\bf \small  Cons} } \\ \midrule
{\small \multirow{4}{*}{\begin{tabular}[c]{@{}l@{}} Customer-\\ focused \\ Heuristics\end{tabular}} } & \begin{tabular}[c]{@{}l@{}} Restricted\\  subgraph   \end{tabular}         &  \begin{tabular}[c]{@{}l@{}} matching with earliest \\ arrival passengers \\awaiting in  subgraph \end{tabular}    &  easy-to-implement     &  large pickup time   \\ \cline{2-5} 
                                                                                        & \begin{tabular}[c]{@{}l@{}} Restricted  subgraph \\greedy method   \end{tabular}          & \begin{tabular}[c]{@{}l@{}}  matching with the \\ nearest driver in\\ subgraph \end{tabular}        &  small pick-up time    & \begin{tabular}[c]{@{}l@{}}   long detour time \\ in the worst case \end{tabular}   \\ \cline{2-5} 
                                                                                        & Bipartite   &   matching with delays          &  \begin{tabular}[c]{@{}l@{}} obtain minimum  \\ total pick-up time  \end{tabular}   & incur holding cost      \\ \hline
{\small \multirow{2}{*}{\begin{tabular}[c]{@{}l@{}}Platform-\\focused \\ Heuristics\end{tabular}} } &  {\small \begin{tabular}[c]{@{}l@{}}Target occupancy with \\ radius \cite{yang2018modelling} \end{tabular} }   &  \begin{tabular}[c]{@{}l@{}}  maintain occupancy \\ by adding   rides \\ in matching radius  \end{tabular}         & industrial practice      &   \begin{tabular}[c]{@{}l@{}}  potential long trip \\ time \end{tabular}  \\ \cline{2-5} 
 & \begin{tabular}[c]{@{}l@{}}Target occupancy with \\ one-sided subgraph  \end{tabular}  &     \begin{tabular}[c]{@{}l@{}}  combination of target \\ and subgraph \\ heuristics \end{tabular}        &  \begin{tabular}[c]{@{}l@{}}  trade-off between \\ pick-up and in- \\vehicle time \end{tabular}  &  $\qquad --$    \\  \hline 
{\small \multirow{2}{*}{Baseline}  }           & {\small \begin{tabular}[c]{@{}l@{}} Path clustering \cite{hong2017commuter} \\ O/D grouping \end{tabular} }      &    matching similar trips    &  \begin{tabular}[c]{@{}l@{}}  near-optimal for \\ offline assignment \end{tabular}    &  \begin{tabular}[c]{@{}l@{}}   unsuitable for \\ real-time assign-\\ment \end{tabular} \\ \cline{2-5}                                                                & {\small \begin{tabular}[c]{@{}l@{}} Single-ride  \cite{schreieck2016matching} \end{tabular} }     &    $\qquad \quad --$    &  $\qquad --$    &  no ride-pooling    \\ \bottomrule
\end{tabulary}
\caption{Summary of heuristics compared in numerical experiments } \label{tab:2}
\end{center}
\end{table*}

\section{Numerical simulation on real-world data} \label{sec4}

\subsection{Simulation environment}
\begin{figure}
    \centering
    \includegraphics[width=\linewidth]{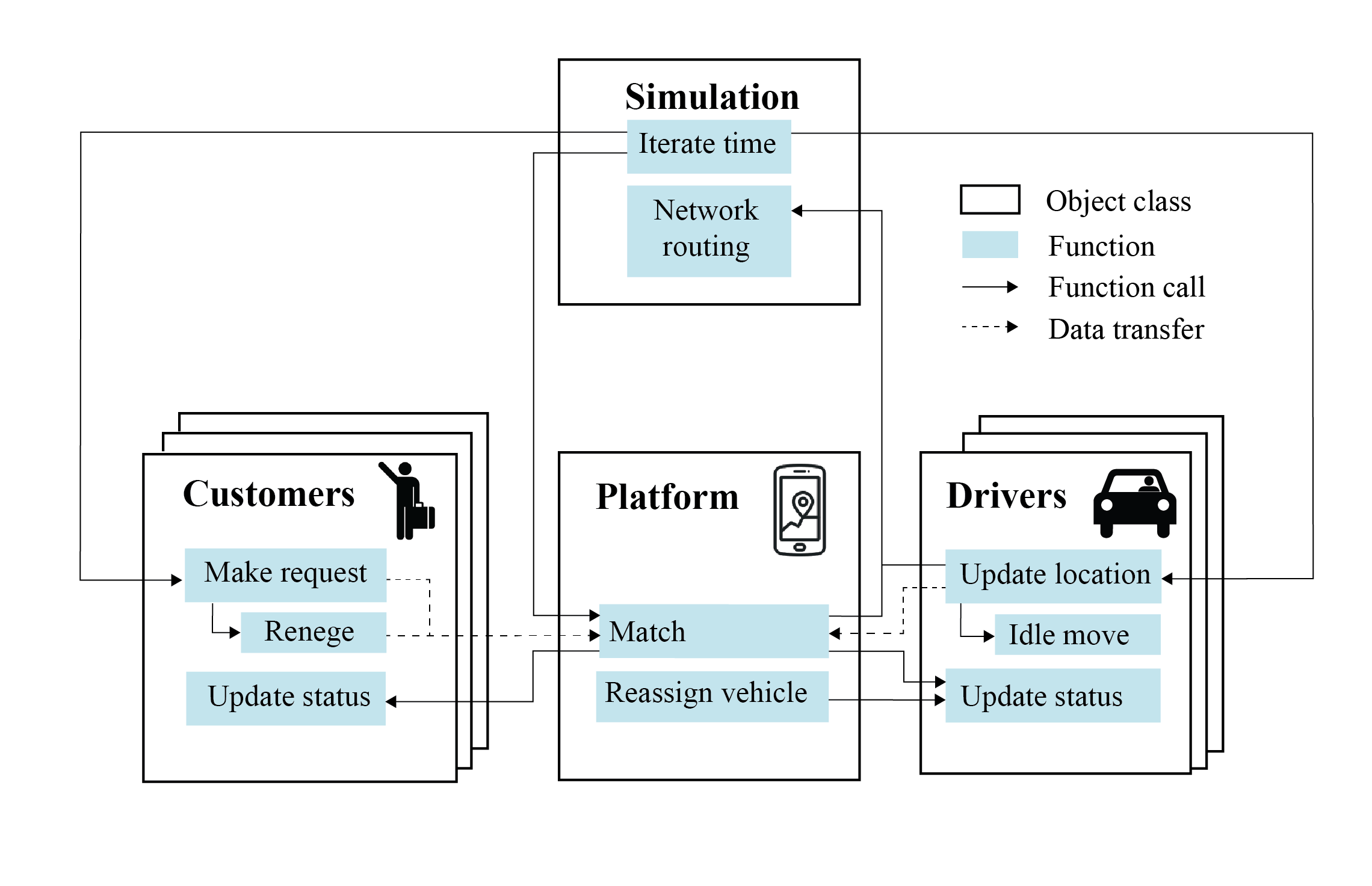}
    \caption{Diagram of relationships between classes in ride-pooling simulator}
    \label{fig:3}
\end{figure}

In order to compare the performance of these heuristics and evaluate their ability to pool requests, we develop an agent-based simulation environment based on a real-world taxicab dataset. Figure \ref{fig:3} is a sketch of the architecture of the simulator. At each timestep, all agents in the simulation are updated, and new trip requests are collected or generated. Drivers are updated first, during which they drop-off or pickup customers if appropriate. Next, waiting customers are updated and determine whether they leave the queue based on their threshold waiting time. Similarly, idle drivers can move randomly or based on given information about supply and demand. Finally, the platform is updated with all the new requests and driver statuses at every matching interval. The platform then performs a new matching assignment and updates customers and drivers with the results. Detailed statistics on miles traveled, driver status and occupancy, and waiting times are updated at each timestep.

This object-oriented simulation environment is designed to be flexible and allows for varied driver, customer, and platform behaviors. In this paper, we assume idle drivers cruise randomly throughout the network when unassigned. However, other behaviors such as navigating towards areas of high demand or platform repositioning are also supported. Additionally, this paper only places constraints on customer matching waiting time behavior and does not investigate different customer pickup waiting time behavior, but this is also supported. The flexibility of this environment (in Python 3.7) allows for potential testing of a multitude of different scenarios with minimal changes. The testing of the various heuristics in this paper is done simply by changing the platform matching function and the randomly sampled demand. Other tests can be done by changing the network, number of drivers, or other parameters easily.

\subsection{Data description}
The data used for the simulation are drawn from multiple sources:
\begin{enumerate}
    \item Trip request data: the NYC Taxicab data set available from the NYC Taxi and Limousine Commission \cite{dias2019fusing}.
    \item Speed data: average street speed data from the Uber movement project \cite{aryandoust2019city}.
    \item Street network data and subgraph network representation: OpenStreetMap road network is converted to a graph in Networkx.
\end{enumerate}

The NYC taxicab data consists of pickup and dropoff latitude and longitude locations for every taxi ride in NYC for 2011-2014. Data from 2013 was used specifically as it is widely used in other papers \cite{Alonso-Mora462, simonetto2019real} and it is the most recent NYC data containing latitude and longitude coordinates instead of more general zone information. In addition, the data contain fare and distance information for the trip, which are not used in the simulation. Since this paper considers low capacity matching for a ridesourcing service, we assume each request to be two or fewer people per request (as is commonly required in TNC pooling services). All heuristics in this paper support larger request sizes. However,  a capacity constraint check would need to be added to make sure that vehicle capacities would not be violated in the matching assignment.


Network average travel times are obtained from the Uber Movement speeds dataset, averaged over 2017-2019 for all links, and synced with the underlying subgraph. Links without any information are given average speeds according to their OpenStreetMap road type (i.e., highway, primary, residential).

\begin{figure*}
    \centering
    \includegraphics[scale=0.6]{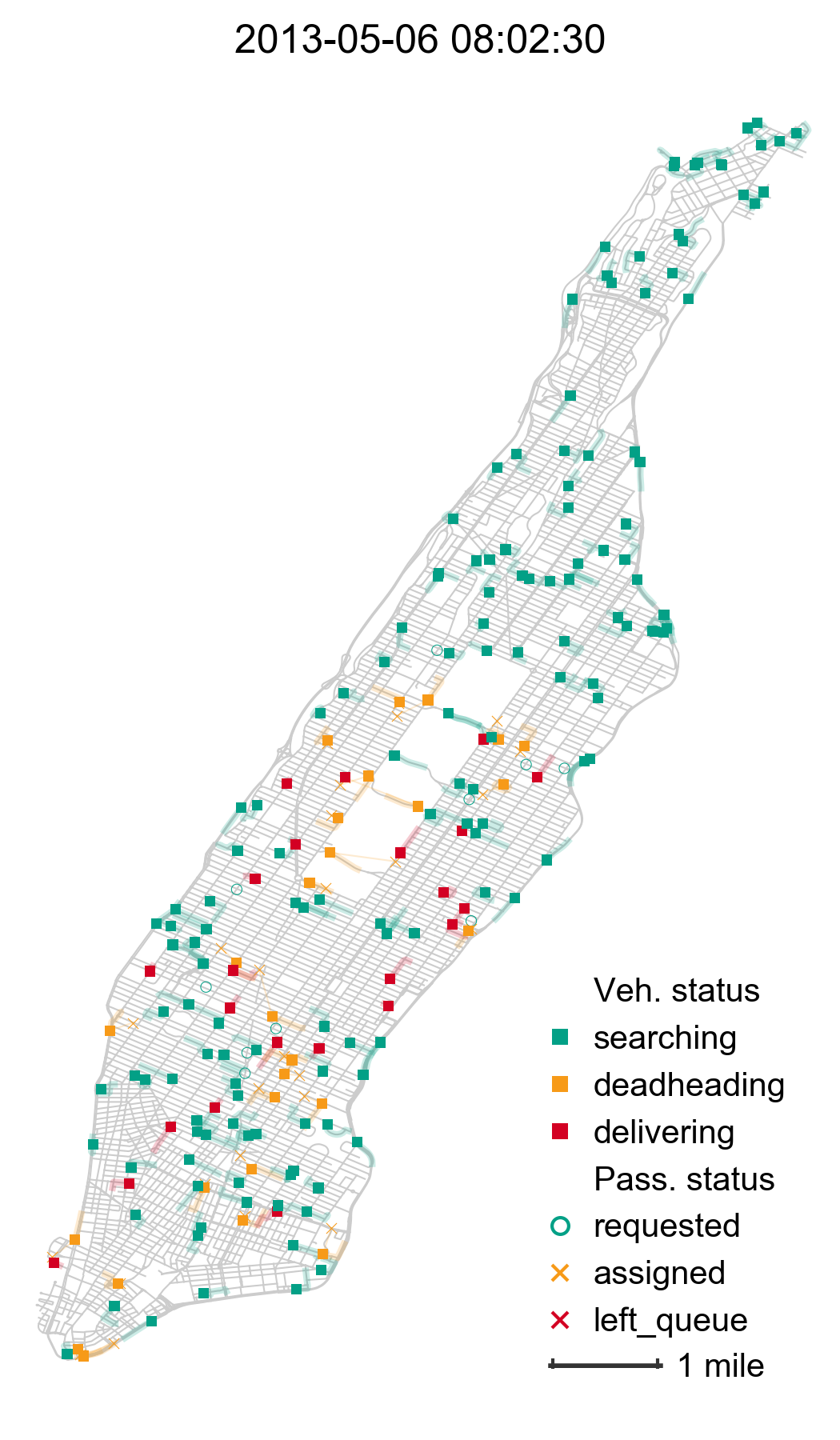}
    \includegraphics[scale=0.6]{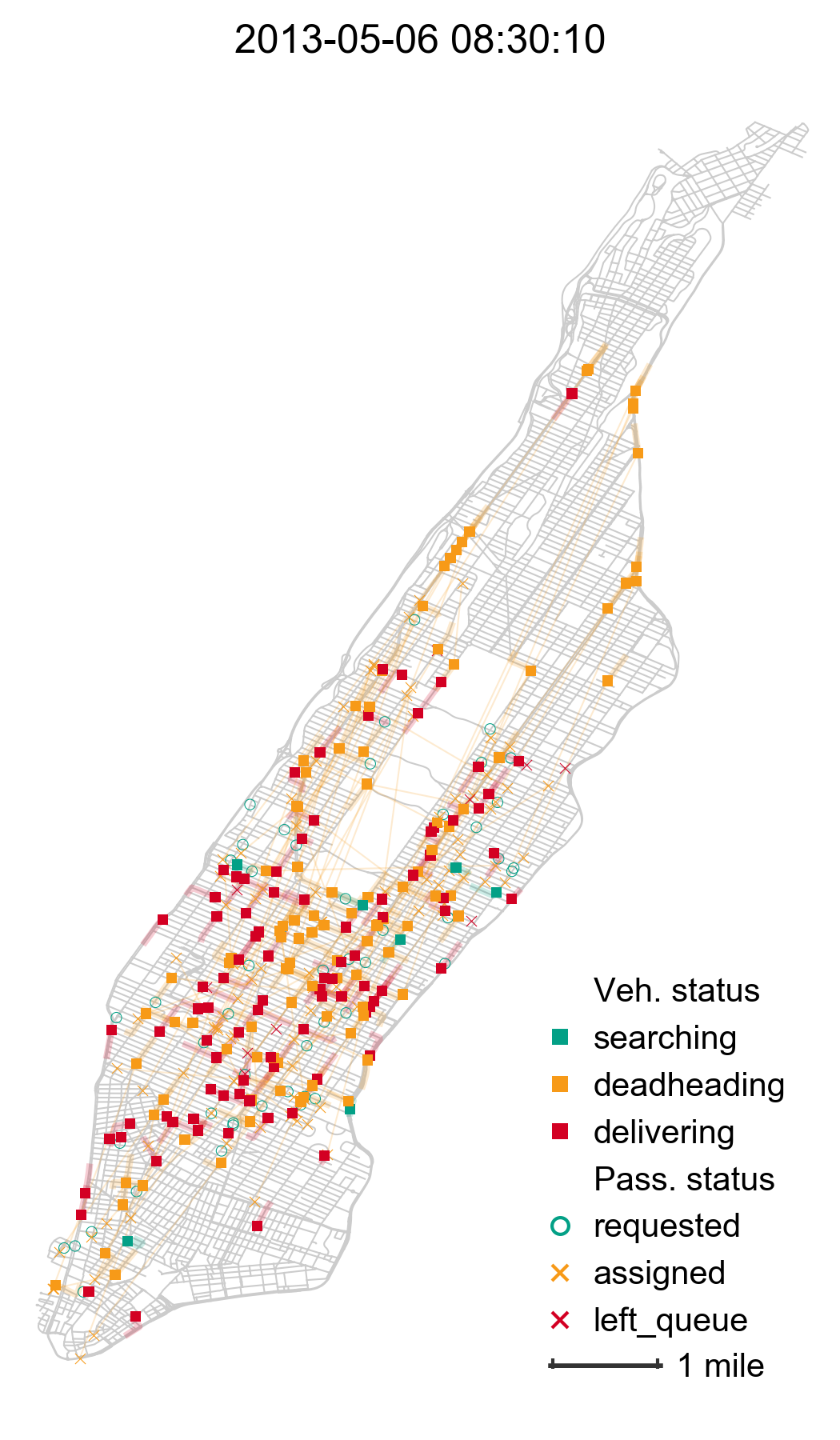}
    \includegraphics[scale=0.6]{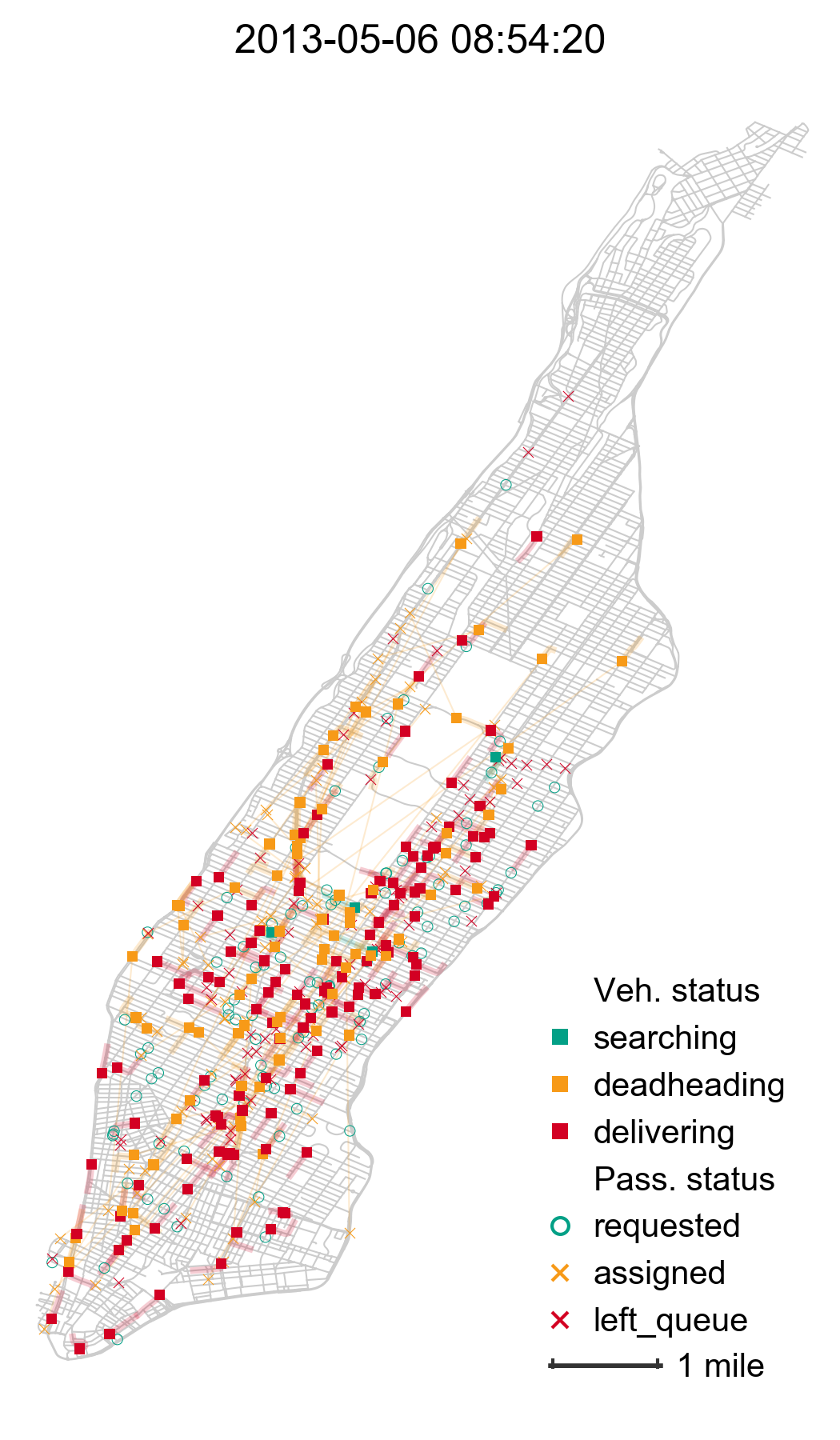}
    \caption{Visualization of the agent-based ridesourcing simulation in Manhattan, New York}
    \label{fig:my_label}
\end{figure*}

\subsection{Results and discussion} 
In order to compare multiple heuristics by statistical tests, we need to specify how to randomly sample supply and demand in the simulator. 

We define a \emph{scenario} as a sampled supply and demand profile and an initial condition for ride-pooling. In each scenario, the demand is randomly sampled from the NYC taxicab data from May 6th, 2013, 7:30 to 9:00 am. In each test, the fleet size is set to 500 vehicles with random initial locations in the road network, and the simulation is run for a warm-up period from 7:30 to 8:00 am. The timestep for updates is every five seconds. The matching interval is also set to five seconds for all heuristics except bipartite and group methods, which use a one-minute interval. Additionally, the target occupancy in the relevant heuristic is two requests per vehicle, and the restricted subgraph methods use an allowable delay function as described earlier. All groups of heuristics are examined with the same initial conditions, but the system evolves differently once trip assignments are made. 

Table \ref{scenarios} summarizes the different scenarios run in simulation. Three different demand cases are examined by sampling the demand at 7\%, 10\%, and 12\% of the total Manhattan demand over the simulation period. Because the fleet size remains constant, these demand levels represent oversupply (enough vehicles to serve all trips with one request per vehicle), nearly balanced  (most strategies can manage the demand), and severe under-supply (where many customers leave before being served, regardless of matching strategy), respectively.

In order to directly compare performance across heuristics, we run ten simulations at each demand level. In each run, a new set of requests are sampled from the taxicab dataset, and new initial vehicle positions are generated. The demand and initial positions are kept the same across the various heuristics so that the runs could be statistically compared.

\begin{table}
\centering
\begin{tabular}{c|c|c|c}
\toprule
\begin{tabular}[c]{@{}c@{}}Percentage \\ of demand \end{tabular} & \begin{tabular}[c]{@{}c@{}}\# of scenarios \\ per heuristic \end{tabular} & \begin{tabular}[c]{@{}c@{}}\# of \\ vehicles \end{tabular} & \begin{tabular}[c]{@{}c@{}}Demand-supply \\ balance \end{tabular} \\ \hline
7\%                  & 10                                                                            & 500                & over-supply           \\ \hline
10\%                 & 10                                                                           & 500                & nearly-balanced   \\ \hline
12\%                 & 10                                                                           & 500                & under-supply   \\ \bottomrule
\end{tabular}
\caption{Summary of simulation scenarios} \label{scenarios}
\end{table}


In accordance with the proposed performance measures, we report the comparisons of heuristics by system and customer metrics.  

\subsection{System metrics}
\begin{figure*}
\begin{center}
\begin{subfigure}{.45\linewidth}
  \centering
  \includegraphics[width=\linewidth]{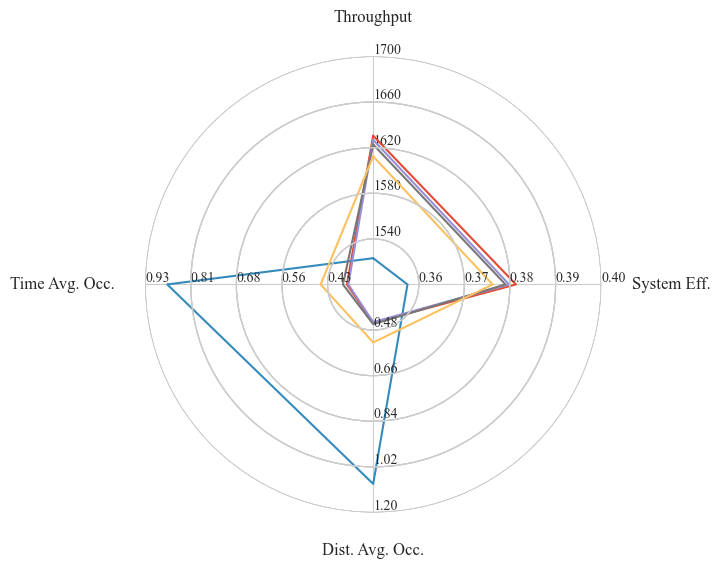}
  \caption{Oversupply (7\% demand)}
  \label{fig:radar7}
\end{subfigure}
    \begin{subfigure}{.45\linewidth}
  \centering
  \includegraphics[width=\linewidth]{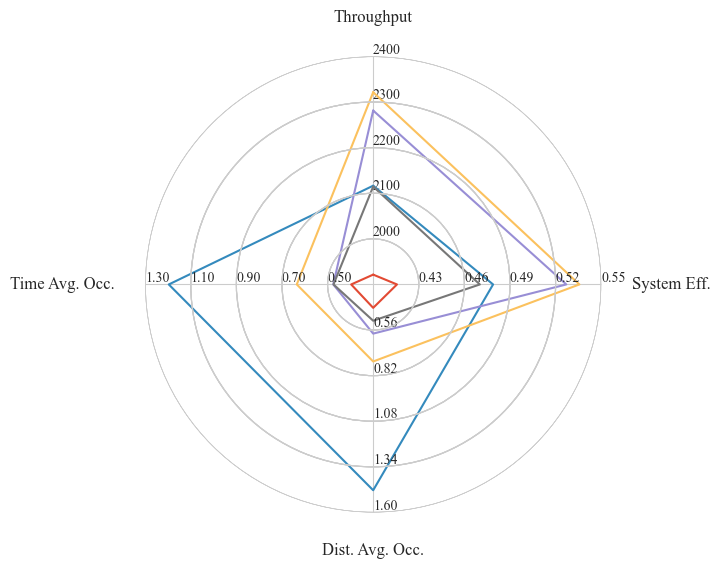}
  \caption{Nearly-balanced (10\% demand)}
  \label{fig:radar10}
\end{subfigure}\\
\begin{subfigure}{.45\linewidth}
  \centering
  \includegraphics[width=\linewidth]{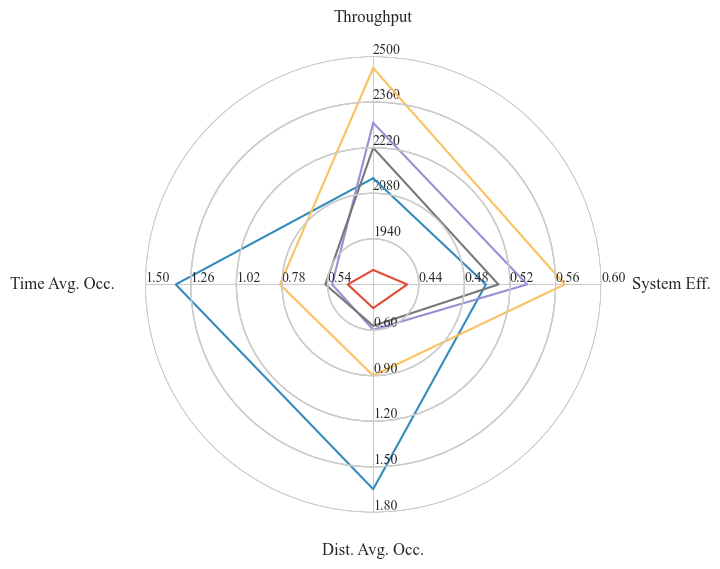}
  \caption{Under-supply (12\% demand)}
  \label{fig:radar12}
\end{subfigure}
\begin{subfigure}{.45\linewidth}
  \centering
  \includegraphics[width=\linewidth]{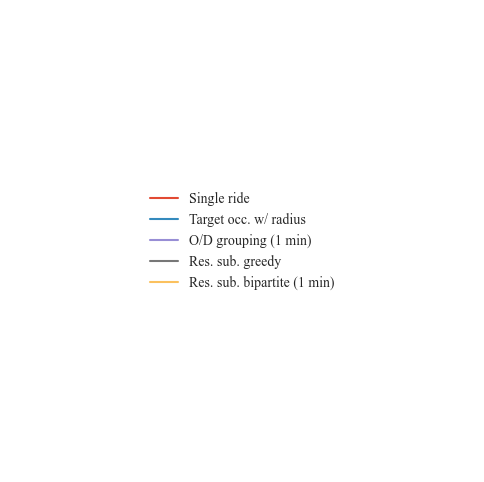}
  \label{fig:radarlegend}
\end{subfigure}
\caption{Summary of results for system metrics}
    \label{fig:radar}
\end{center}
\end{figure*}

We examine the system performance metrics of these methods in closer detail in those three scenarios. These metrics are plotted for comparison in the radar charts in Figure \ref{fig:radar}.

In the oversupply scenario (7\% demand), because all requests can be served without ride-pooling, the single ride and greedy restricted subgraph methods outperform the others in both throughput and system efficiency, with no significant difference between them. As seen in Figure \ref{fig:radar7}, there is a less than 5\% difference between the throughput of other methods except for the target occupancy method. The target occupancy method performs quite poorly while maintaining high occupancy because it may cause long detours by forcing rides to share and low throughput. 

In the nearly-balanced scenario (10\% demand), a much larger variation can be observed between the methods. O/D grouping and bipartite methods, due to their longer matching window or centralized optimization, outperform the other methods in throughput by about 10\% (200 requests per hour). However, this is a trade-off with the additional matching time incurred. Shorter matching windows yield lower overall throughput due to less time to group or consider requests for optimality. It is also important to note that the performance of the O/D grouping method is highly dependent on the spatial distribution of demand. It is very effective only when many requests share similar origins and destinations.

As vehicle supply becomes exceptionally constrained in the 12\% demand scenario, the advantage of the restricted subgraph method and its extensions becomes clearer. Though statistically significant, the O/D grouping platform's average throughput is only ten requests per hour more (or less than 1\%). Meanwhile, the subgraph method depends less on the distribution of demand and has a lower matching time. In most cases, we observe a correlation between throughput and efficiency, as higher efficiency serves more customers with the available labor supply. However, we note that the target occupancy is a consistent exception to this, as it achieves comparable efficiency to some other methods by getting people into vehicles as quickly as possible, but often suffers in throughput due to long detours. This result demonstrates the danger of comparing performance based on only one metric.

\subsection{Customer metrics}
While throughput and efficiency are essential to the platform and regulators, they do not have a direct impact on customers. Thus it is also important to examine metrics that affect customers' experience. There is a clear trade-off between customer satisfaction and platform goals: longer matching time intervals for holding and optimizing requests produce larger throughput and more efficient matches but at the cost of customers' matching time. We also see the effects of different strategy choices. For example, the target occupancy method achieves low pickup times in general due to its matching radius requirement. However, it also experiences long delays because the trip assignments do not take destinations into account.

\begin{figure}
\centering
\includegraphics[width=\linewidth]{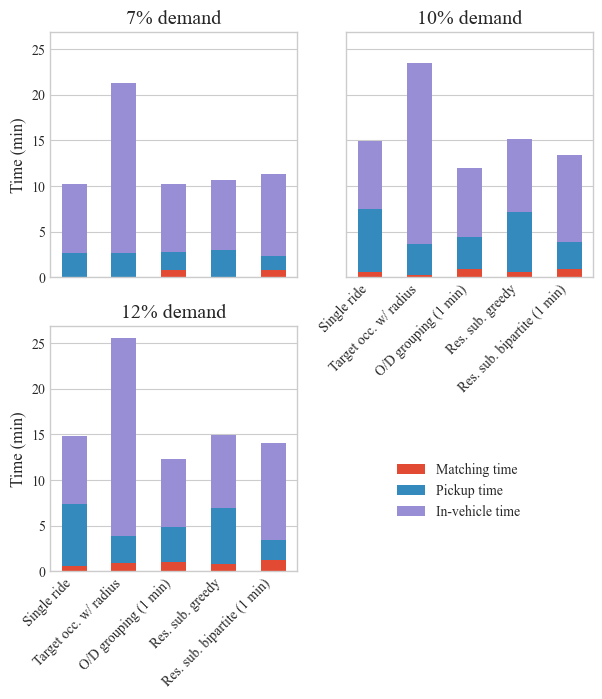}

\caption{Summary of results for customer metrics}
\label{fig:bar}
\end{figure}

\begin{table*}
\centering
\begin{tabular}{c|c|c|c|c|c} 
\toprule
\multicolumn{6}{c}{ Over-supply case with 7\% demand}  \\ \midrule
Platform &
\begin{tabular}[c]{@{}c@{}} Average \\ matching time\end{tabular} & \begin{tabular}[c]{@{}c@{}} Average \\  pickup time \end{tabular} &
\begin{tabular}[c]{@{}c@{}} Average \\  trip time \end{tabular}   &
\begin{tabular}[c]{@{}c@{}} Average total \\ wait time \end{tabular} &
\begin{tabular}[c]{@{}c@{}} Average \\  delay time \end{tabular} \\ \hline
Single ride   & 0.07              & 2.75        & 7.55 & 2.82   & 0.74           \\ \hline
\begin{tabular}[c]{@{}c@{}}Target occupancy\\ with radius\end{tabular} & 0.07              & 2.78            & 18.60  & 2.85 & 11.64          \\ \hline
\begin{tabular}[c]{@{}c@{}}Restricted subgraph\\ greedy method\end{tabular} & 0.07              & 2.69    & 7.61 & 2.76          & 0.80           \\ \hline
\begin{tabular}[c]{@{}c@{}}Res. sub bipartite \\ matching (1 min)\end{tabular}  & 0.75              & 2.57  & 9.13 & 3.32          & 2.35           \\ \hline
O/D grouping (1 min)  & 0.75              & 3.73    & 7.66 &     4.48     & 0.82           \\ \midrule
\multicolumn{6}{c}{ Nearly-balanced case with 10\% demand}  \\ \midrule
Single ride & 0.52              & 8.05            & 7.51          & 8.56                & 0.74           \\ \hline
\begin{tabular}[c]{@{}c@{}}Target occupancy\\ with radius\end{tabular} & 0.25              & 3.59            & 19.70         & 3.84                & 12.77          \\ \hline
\begin{tabular}[c]{@{}c@{}}Restricted subgraph\\ greedy method\end{tabular} & 0.58              & 7.51            & 8.01          & 8.09                & 1.34           \\ \hline
\begin{tabular}[c]{@{}c@{}}Res. sub bipartite \\ matching (1 min)\end{tabular}  & 0.90              & 3.74            & 9.88          & 4.64                & 3.16           \\ \hline
O/D grouping (1 min)  & 0.86              & 6.09            & 7.64          & 6.95                & 0.83           \\ 
\midrule
\multicolumn{6}{c}{ Under-supply case with 12\% demand}  \\ \midrule
Single ride                                                      & 0.52              & 7.96            & 7.49          & 8.48                & 0.74           \\ \hline
\begin{tabular}[c]{@{}c@{}}Target occupancy\\ with radius\end{tabular}                                                   & 0.82              & 3.79            & 21.64         & 4.60                & 14.72          \\ \hline
\begin{tabular}[c]{@{}c@{}}Restricted subgraph\\ greedy method\end{tabular} & 0.76              & 7.40            & 8.18          & 8.17                & 1.53           \\ \hline
\begin{tabular}[c]{@{}c@{}}Res. sub bipartite \\ matching (1 min)\end{tabular}  & 1.24              & 3.43            & 10.92         & 4.68                & 4.30           \\ \hline
O/D grouping (1 min)   & 1.01              & 8.06            & 7.66          & 9.07                & 0.83           \\ \bottomrule
\end{tabular}
\caption{Summary of customer metric results, in minutes, for varying demand levels } \label{tab:cust}
\end{table*}

In the oversupply scenario (7\% demand) in Table \ref{tab:cust}, the matching time is significantly increased for the bipartite matching and O/D grouping algorithms, as they employ a longer matching interval. This is disadvantageous to the consumers because they are forced to wait longer in uncertainty (i.e., they do not know when they will be matched), with often more extended pickup or trip times. We also see that the greedy restricted subgraph algorithm reduces the pickup time as intended in this case.

In the nearly-balanced scenario (10\% demand) in Table \ref{tab:cust}, we see a significant reduction in pickup time using either the target occupancy and the bipartite matching method compared to the other platforms. This reduction comes at the cost of increased matching time, which customers might place a great value on, given the uncertainty of whether they will be assigned a vehicle at all. 

Finally, the waiting and delay times increase further as supply is more constrained and more requests are served in the 12\% demand scenario. In Table \ref{tab:cust}, we see the same trends in average matching time and pickup time, as well as increased trip delays in the target occupancy and bipartite methods. As more demand is trying to be served by the same amount of vehicles, the increased throughput comes with forcing longer delays. We present the restricted subgraph methods as a balance in these trade-offs between the platform's goals and customers' satisfaction, though TNCs and planners can determine what is best for specific scenarios by weighing more important metrics. 

\subsection{Statistical test results for heuristics}

\begin{figure}
    \centering
    \includegraphics[width=\linewidth]{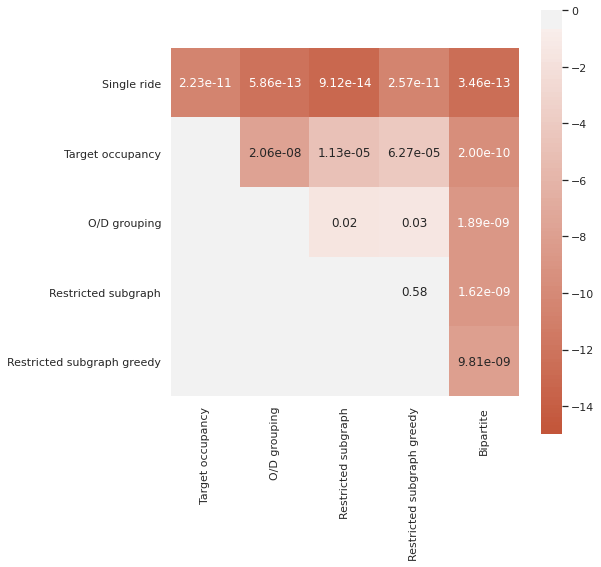}
    \caption{P-value of pairwise t-test. The smaller value indicates that the difference of system throughput using two algorithms is significant. The colorbar is in logarithm scale. }
    \label{fig:p-value}
\end{figure}

We perform a pairwise t-test to compare throughput performance based on ten simulation runs at each demand level. The null hypothesis, $h_0$, is that the methods do not significantly differ in the system throughput. The throughput and p-values for the 12\% demand scenario comparisons are reported in Figure \ref{fig:p-value}. We only present the 12\% demand statistical comparison here for brevity. The tests show statistical significance in most demand levels in all pairs except between the restricted subgraph method and the greedy restricted subgraph, which only have a statistically significant difference in the 7\% demand scenario. It is important to note that, while these differences are often statistically significant, they are not always practically significant. As mentioned earlier, in some cases, the difference between O/D grouping and restricted subgraph methods is less than one percent.

\subsection{Results takeaways}
In general, our results show that:
\begin{itemize}
    \item The restricted subgraph method performs significantly better than the target occupancy method, a method practiced in industry, in all cases due to increased throughput and limited detour time. 
    \item  The performance of the O/D grouping method, another one practiced in industry, depends on tuned parameters (e.g., the size of matching zones and the length of matching window). In under-supply scenarios with a long matching window, the restricted subgraph is on par with the throughput produced by the O/D grouping method while having shorter average matching times. In under-supply cases but with shorter matching windows, the restricted subgraph method achieves about 3\% better throughput without needing adjustment. The only parameter in the subgraph method is the delay function dictated by perceived customer preference.
    \item The restricted subgraph method performs typically within 10\% of the throughput of the bipartite matching method in under-supply, with lower matching times and lower trip times. Bipartite matching aiming to improve the matching reward inversely stretches people to the limits of their delays. Besides, the computing challenge for the near-optimal bipartite optimization grows exponentially in dense areas.
    \item Greediness in trip assignments improves the fleet utilization in oversupply scenarios by reducing average pickup times. 
    \end{itemize}

In summary, the simulation results demonstrate that the proposed restricted subgraph heuristic can well balance the interests of platforms and passengers. It limits their trip delay and improves their experience without compromising the system's productivity and efficiency. 


\section{Conclusion} \label{sec5}
In order for the ridesourcing industry to continue to grow, it must acknowledge and work to limit the impact it has on traffic congestion in dense urban areas. An important aspect of this will be the use and encouragement of ride-pooling to serve the same number of trips with fewer vehicles and fewer miles travelled. The restrict subgraph heuristic for ride-pooling assignment proposed in this work is a robust and easily implementable method as a substitute for the centralized optimization methods. It can achieve modest performance when facing a severe scalability issue in dense urban areas. 

Future investigations are necessary to address the theoretical foundation of the ride-pooling heuristic methods.  In addition, further development on a ride-pooling metaheuristic, i.e., a procedure to select suitable heuristics, is a promising direction. As seen in the numerical results, conditioning on the supply-demand relationship in ridesourcing platforms, the discussed heuristics and baseline single-ride algorithms all have their advantages and disadvantages. Developing a data-driven method that automatically chooses the appropriate methods is a meaningful supplement to the literature.

\section{Acknowledgements}
The work described in this paper was partly supported by research grants from the Ford Motor Company and the National Science Foundation (CMMI-1740865; 1854684). 

\newpage

\bibliographystyle{TRR}
\bibliography{main}
\end{document}